\documentclass[10pt,twocolumn,twoside]{IEEEtran}

\usepackage[bookmarks=false,hyperfigures=false,hidelinks=true]{hyperref}

\usepackage{url, doi}

\usepackage{amsmath}
\usepackage{amssymb, subcaption, xspace, diagbox}
\usepackage{mathtools}

\usepackage{algorithm2e}
\providecommand{\dontprintsemicolon}{\DontPrintSemicolon}

\usepackage[final]{graphicx}
\DeclareGraphicsExtensions{.pdf,.png}
\usepackage{color}

\usepackage{tikz, pgfplots}
\pgfplotsset{compat=1.15}

\newcommand{\mgt}[1]{\textbf{#1}}
\newcommand{\gt}[1]{(#1)}
\newcommand{\lt}[1]{{#1}}

\newcommand{\ie}{i.e.\xspace}
\newcommand{\eg}{e.g.\xspace}

\newcommand{\eq}[1]{(\ref{eq:#1})}
\newcommand{\fig}[1]{Fig.~\ref{fig:#1}}
\newcommand{\figs}[1]{Figs.~\ref{fig:#1}}
\newcommand{\fign}[1]{\ref{fig:#1}}
\newcommand{\tbl}[1]{Table~\ref{tbl:#1}}
\newcommand{\sctn}[1]{Sec.~\ref{sec:#1}}
\newcommand{\appx}[1]{Appx.~\ref{sec:#1}}

\newcommand{\mb}[1]{\mathbf{#1}}

\newcommand{\mbb}[1]{\mathbb{#1}}
\newcommand{\mc}[1]{\mathcal{#1}}

\newcommand{\prox}{{\mathrm{prox}}}

\newcommand{\norm}[1]{\left\| #1 \right\|}
\DeclarePairedDelimiterX{\normsz}[1]{\lVert}{\rVert}{#1}

\DeclareMathOperator*{\argmin}{arg\,min}
\DeclareMathOperator*{\argmax}{arg\,max}

\newcommand{\sinc}{\mathop{\mathrm{sinc}}}

\date{\today}

\begin{document}

\title{PSF Estimation in Crowded Astronomical Imagery as a Convolutional Dictionary Learning Problem}

\author{Brendt Wohlberg and Przemek Wozniak%
\thanks{B. Wohlberg is with Theoretical Division, Los Alamos
    National Laboratory, Los Alamos, NM 87545, USA (Email:
    \texttt{brendt@lanl.gov})}
\thanks{P. Wozniak is with Space and Remote Sensing Group,
    Los Alamos National Laboratory, Los Alamos, NM 87545, USA (Email:
    \texttt{wozniak@lanl.gov})}
\thanks{Research presented in this article was supported by the Laboratory Directed Research and Development program of Los Alamos National Laboratory under project numbers 20170183ER and 20200061DR.}
}

\maketitle

\begin{abstract}
We present a new algorithm for estimating the Point Spread Function (PSF) in wide-field astronomical images with extreme source crowding. Robust and accurate PSF estimation in crowded astronomical images dramatically improves the fidelity of astrometric and photometric measurements extracted from wide-field sky monitoring imagery. Our radically new approach utilizes convolutional sparse representations to model the continuous functions involved in the image formation. This approach avoids the need to detect and precisely localize individual point sources that is shared by existing methods. In experiments involving simulated astronomical imagery, it significantly outperforms the recent alternative method with which it is compared.
\end{abstract}


\noindent \emph{This is an extended version of an IEEE Signal Processing Letters paper (\doi{10.1109/LSP.2021.3050706}), with supplemental material included as appendices.}

\IEEEpeerreviewmaketitle

\section{Introduction}
\label{sec:intro}

Astronomical images deliver a wealth of information on a wide range of phenomena in natural objects such as stars and galaxies. Similar techniques have been successfully applied to tracking man made space objects, showing great promise to address pressing problems in Space Traffic Management~\cite{bhavya-2018-global}. Point Spread Function (PSF) estimation in astronomical imagery presents unique challenges~\cite{racine-1996-telescope-psf, lupton-2007-astro-imaging}. Stars are nearly perfect point sources, so there is no shortage of fiducial points for analysis. At the same time, there are numerous factors that affect the PSF shape: atmospheric blur, imperfect optics and sky tracking, vibration etc.  Modeling is often performed iteratively, using stars to improve the PSF model and using the model to better fit all stars~\cite{schechter-1993-dophot}.  While deconvolution is often considered more fundamental in signal processing~\cite{campisi-2017-blind, chaudhuri-2014-blind}, many applications in astronomy are framed as PSF estimation and forward modeling in the convolved image~\cite{lupton-2007-astro-imaging, mandelbaum-2018-weak-lensing}.  For example, changes in brightness and motion of unresolved sources are typically extracted by fitting individual PSF profiles and "streaks" in differenced (uncluttered) images of the same field separated in time~\cite{wozniak-2018-moving}. PSF fitting on the original crowded images is performed to measure positions and brightness of stars used for calibration and science. Reaching the required fidelity is rarely possible without a good subpixel PSF model.

The analysis of crowded stellar fields is an important and challenging application of astronomical imaging~\cite{wozniak-2008-crowded-fields}. When deep source confusion sets in, every image pixel includes signal from multiple PSF profiles. This situation naturally arises in densely populated sky areas and in very wide-field imaging that aims to cover as many objects as possible~\cite{piotrowski-2013-wide-field}. There is a scarcity of algorithms and software tools that can tackle extreme crowding. Standard source extraction and PSF estimation codes like DAOPHOT~\cite{stetson-1987-daophot}, DoPHOT~\cite{schechter-1993-dophot}, SExtractor~\cite{bertin-1996-sextractor} were not designed to handle images where not a single star can be considered sufficiently isolated to ignore perturbations from neighbors. Their treatment of crowding typically consists of identifying occasional PSF collisions to either fit special local models or eliminate them from consideration.
Recent PSF estimation work in astronomy has focused primarily on super-resolution and sub-pixel sampling by paying close attention to the correct image formation model and introducing modern sparsity based approaches (e.g.~\cite{anderson_2000-toward, ngole-2014-superresolution, ngole-2016-constraint}). These algorithms are an important step forward, but they still ignore the cross-talk between sources and rely on user's ability to identify isolated stars.

In this paper we present a new PSF estimation algorithm based on convolutional sparse representation (CSR). There is no need to detect and fit individual stars, eliminating the uncertainties and instabilities associated with these local modeling decisions.  We are not aware of any prior use of CSR methods for this application.\footnote{These methods have previously been considered for analysis of astronomical imagery~\cite{pla-2018-convolutional}, but the application was background removal rather than PSF estimation, and no attempt was made to model the continuous nature of the underlying scene.}

We also note that the methods presented here include some more general contributions in CSR, including the use of an interpolation kernel to generate a dictionary suitable for approximating the translations of a continuous function, as well as additional algorithm refinements described in~\sctn{cdl}. Other authors have also devised techniques for CSR of continuous signals~\cite{ekanadham-2011-recovery, tang-2013-sparse, soh-2020-group}, but employing very different methods. The approach of~\cite{song-2020-convolutional}, which we became aware of during the final stages of preparation of this manuscript, exploits similar ideas to ours in the use of interpolation to generate the dictionary, but makes use of greedy algorithms as opposed to our optimization-based approach. The latter has the advantage of greater flexibility supporting different regularization terms and constraints, which is exploited in constructing our proposed PSF estimation method.

\section{Image Formation Model}
\label{sec:model}

We restrict our attention to estimation of a spatially-invariant PSF. In practice it is usually necessary to characterize imaging systems with a spatially-varying PSF, but since these variations are typically negligible across the small image regions required by our approach, they can be represented by making independent estimates of a fixed PSF in overlapping image regions covering the image. We represent the scene being imaged as the continuous function $r(x,y)$, where $x$ and $y$ are spatial coordinates, the image on the detector as the continuous function $s(x,y)$, and the PSF of the optical system by the continuous function $g(x,y)$, so that we have (ignoring noise for now)
\begin{equation}
s(x, y) = \int_{-\infty}^{\infty} \int_{-\infty}^{\infty} r(x - u, y - v)
          g(u, v) \,du \,dv \,.
\label{eq:conv}
\end{equation}
In the case of an ideal detector, the final sampled version of the image, $\mb{s}$, is obtained by point sampling of the image function $s(x, y)$. In practice, however, detectors sample the image function $s(x, y)$ by integrating its product with some sensitivity function at each photosite. This behavior can be modeled as the convolution of $s(x, y)$ by the photosite sensitivity function, followed by point sampling. As a result of the commutative property of convolution, this additional convolution can be included in~\eq{conv} by redefining $s(x,y)$ as the image on the sensor blurred by the photosite sensitivity function, and $g(x,y)$ as the convolution of the PSF of the optical system and the photosite sensitivity function. It is this \emph{effective} PSF~\cite{anderson_2000-toward} that we will be estimating.

Our image formation model assumes that the scene consists of a finite sum of impulses
\begin{equation}
r(x, y) = \sum_k a_k \delta(x - x_k, y - y_k) \;,
\label{eq:scene}
\end{equation}
where $a_k$, $x_k$, and $y_k$ are the scaling factor and $x$ and $y$ locations respectively of the $k^{\text{th}}$ impulse, so that we have
\begin{align}
s(x, y) = & \iint \sum_k a_k
          \delta(x - x_k - u, y - y_k - v) g(u, v) \,du \,dv \nonumber \\
        = & \sum_k a_k g(x - x_k, y - y_k) \;.
\label{eq:contmod}
\end{align}
If the $x_k$, and $y_k$ values were quantized to a finite resolution grid, this equation could be equivalently represented in discrete form as $\mb{s} = \mb{g} \ast \mb{a} \;,$ where $\mb{s}$ and $\mb{g}$ denote $s(\cdot,\cdot)$ and $g(\cdot,\cdot)$ sampled on that grid, and $\mb{a}$ is an image, on the same sampling grid, taking on the value zero except at sample positions corresponding to one of the $x_k, y_k$ pairs above.


\def\gauss(#1,#2){1/(#2*sqrt(2*pi))*exp(-((#1)^2)/(2*#2^2))}%
\pgfmathsetmacro{\sgma}{0.65}
\pgfmathsetmacro{\xa}{3.6}
\pgfmathsetmacro{\xb}{9}
\def \gsum(#1){\gauss(#1 - \xa, \sgma) + \gauss(#1 - \xb, \sgma)}

\begin{figure}[htbp]
\begin{tikzpicture}
\begin{axis}[
  no markers, domain=0:12, samples=200,
  axis lines=left, xlabel=$x$, ylabel=$y$,
  every axis y label/.style={at=(current axis.above origin),anchor=south},
  every axis x label/.style={at=(current axis.right of origin),anchor=west},
  height=4.5cm, width=9.8cm,
  xtick=\empty, ytick=\empty,
  enlargelimits=false, clip=false, axis on top,
  grid = major
  ]
  \addplot [very thick,cyan!50!black] {\gsum(x)};
  \pgfplotsinvokeforeach{0,...,11}{
    \draw [fill=black](axis cs:#1, {\gsum(#1)}) circle [radius=0.65mm];
    \draw [thick,black] (axis cs:#1,0) -- (axis cs:#1, {\gsum(#1)});
  \draw [loosely dotted,thick,red!75!black] (axis cs:3.6,0) -- (axis cs:3.6,0.61);
  \draw [loosely dotted,thick,red!75!black] (axis cs:9,0) -- (axis cs:9,0.61);
}
\end{axis}
\end{tikzpicture}
\caption{Illustration of dependence of sampled PSF values on alignment of the PSF with the sampling grid. The dotted red lines indicate the location of the impulses defining the locations of the PSFs. \vspace{-1.5mm}}
\label{fig:psfsample}
\end{figure}
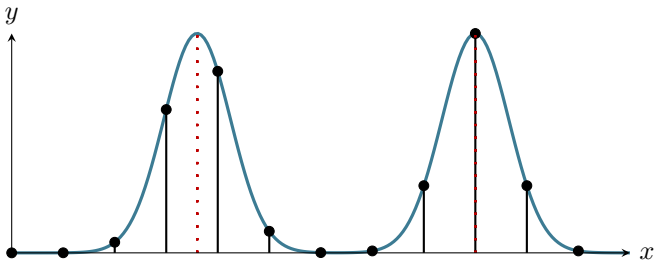


In this simplified context, a natural approach to the PSF estimation problem would be to exploit the sparsity of $\mb{a}$, posing the problem as blind deconvolution via regularized inversion with a sparsity prior, \eg
\vspace{-1mm}
\begin{equation}
  \argmin_{\mb{g}, \mb{a}} \; (1/2) \norm{ \mb{g} \ast \mb{a} - \mb{s} }_2^2 + \lambda \norm{\mb{a}}_1
\label{eq:blnddcnv}
\end{equation}
with a squared $\ell_2$ data fidelity term\footnote{The Poisson noise model encountered in practice suggests that we should at least employ an appropriate weighted $\ell_2$ data fidelity term~\cite[Ch. 17]{bouman-2020-model}. We retain the unweighted norm since the Poisson noise weighting was found to complicate algorithm convergence without providing any significant performance improvements.} and an $\ell_1$ regularization term. However, since our images are typically sampled close to the Nyquist rate, different alignments of the signal with respect to the sampling grid can result in significant differences in the samples obtained from the same continuous signal, as illustrated in~\fig{psfsample}.

\section{Convolutional Dictionary Learning}
\label{sec:cdl}

In this section, for simplicity of notation, concepts are introduced and mathematically defined in the context of 1D signals. The extension to the 2D signals is, for the most part trivial, and details of the extension are explicitly provided when it is not.  While the simple convolutional model $\mb{s} = \mb{g} \ast \mb{a}$ is not entirely adequate, a significantly more accurate discrete model can be defined as $\mb{s} = \sum_m \mb{g}_m \ast \mb{a}_m \,,$ where $\mb{g}_m$ denote different sub-pixel sampling offsets of the continuous function $g(\cdot)$, and the $\mb{a}_m$ are corresponding maps of the sub-pixel impulse locations as in~\eq{contmod}. A naive extension of~\eq{blnddcnv} to account for this model would be
\begin{equation}
  \argmin_{\{\mb{g}_m\}, \{\mb{a}_m\}} \; \frac{1}{2} \normsz[\Big]{ \sum_m \mb{g}_m \ast \mb{a}_m - \mb{s} }_2^2 + \lambda \sum_m \norm{\mb{a}_m}_1 \;,
\label{eq:cdl}
\end{equation}
\ie a convolutional dictionary learning (CDL) problem~\cite{garcia-2018-convolutional}.

We modify the generic CDL problem for our purposes by defining the $\mb{g}_m$, sampled at different sub-pixel offsets, to be derived via linear interpolation from a common grid-aligned (\ie zero sub-pixel offset) PSF kernel $\mb{g}$. Since linear interpolation to a set of $M$ fractional offsets from the sampling grid can be computed via convolution with a set of $M$ filters $\{ \mb{h}_m \}$, we can write dictionary filters  $\mb{g}_m$ as\footnote{In two dimensions we have $M^2$ filters $\mb{g}_{m, n} = (\mb{h}_m \otimes \mb{h}_n) \ast \mb{g}$, where $m$ and $n$ index the fractional offsets on the two axes, and $\otimes$ denotes the tensor product of two vectors.}
\begin{equation}
   \mb{g}_m =  \mb{h}_m \ast \mb{g} \;.
\end{equation}
We use Lanczos interpolation~\cite[Sec. 10.3.6]{burger-2009-principles2}, for which the interpolation kernel of order $K$ is defined as
\[
\phi(x) = \left\{ \begin{array}{ll}
                  \sinc(x) \sinc(x / K) & \text{if} \;\; -K < x < K \\
                  0 & \text{otherwise}
                   \end{array} \right.\,\;,
\]
where $\sinc(x) = sin(\pi x) / (\pi x).$ Defining the set of fractional offsets (chosen to evenly divide the intervals between the integer grid points) as values $n / M$ where $n \in \mbb{Z}$ and $ -\lfloor (M - 1) / 2 \rfloor \leq n \leq \lfloor M / 2 \rfloor$,  filter $\mb{h}_m$ is obtained by evaluating $\phi(x)$ at the set of points
 $\{-K + \delta_m, -K + 1 + \delta_m, \ldots, K - 1 + \delta_m, K + \delta_m \}$, where $\delta_m$ is the $m^{\text{th}}$ fractional offset.

We can therefore pose our variant of the CDL problem as
\begin{equation}
  \argmin_{\mb{g}, \{\mb{a}_m\}} \; \frac{1}{2} \normsz[\Big]{ \sum_m \mb{h}_m \ast \mb{g} \ast \mb{a}_m - \mb{s} }_2^2 + \lambda \sum_m \norm{\mb{a}_m}_1 \;,
\label{eq:cdlzs}
\end{equation}
which, as usual for such bi-convex problems, is solved via alternating minimization with respect to the $\mb{a}_m$ and $\mb{g}$. By associativity of convolution we can express the minimization with respect to the $\mb{a}_m$ as a
convolutional sparse coding (CSC)~\cite{wohlberg-2016-efficient} of $\mb{s}$ with respect to dictionary $\mb{d}_m = \mb{h}_m \ast \mb{g}$, and by commutativity and linearity of convolution we can express the minimization with respect to $\mb{g}$ as a deconvolution of $\mb{s}$ with respect to the kernel $\mb{b} = \sum_m \mb{h}_m \ast \mb{a}_m$.

We introduce a number of additional refinements for improved performance:
\subsubsection*{DC invariance}
Astronomical imagery includes a very smooth background that can be accurately modeled as a constant offset on spatial scales of up to a few hundred pixels. In practice, this amounts to a non-zero DC offset that is omitted from image formation model~\eq{scene}, and is not accounted for in the data fidelity term of our CDL problem, resulting in poor performance due to the mismatch between the model and the data. The most effective solution is to include a frequency-domain mask in the data fidelity term that excludes the DC value from having any effect. This is straightforward to implement since both the sparse coding and dictionary update sub-problems employ frequency-domain solvers~\cite{garcia-2018-convolutional}.
\subsubsection*{Non-negativity}
Both $\mb{g}$ and the $\mb{a}_m$ must be non-negative according to the physical process being modeled. This requirement is included as an additional constraint on $\mb{g}$, but is omitted for $\mb{a}_m$ since it was empirically observed not to make a significant performance difference.
\subsubsection*{Normalization of $\mb{g}$}
We include a unit-norm constraint on $\mb{g}$ to resolve the scaling ambiguity between $\mb{g}$ and the $\mb{a}_m$.
\subsubsection*{Regularization of the $\mb{a}_m$}
In the CSC sub-problem, we replace the usual $\ell_1$ norm regularizer with an $\ell_1 \!-\! \ell_2$ norm, which has been demonstrated to provide improved sparse recovery with a highly coherent dictionary~\cite{lou-2015-computing}.
\subsubsection*{Regularization of $\mb{g}$}
We include a regularization term consisting of the squared $\ell_2$ norm of the gradient of $\mb{g}$~\cite[Sec. 4]{wohlberg-2016-convolutional2}, which has the effect of penalizing non-smooth solutions.

The resulting CDL problem can be written as
\begin{align}
 \argmin_{\mb{g}, \{\mb{a}_m\}} \; & \frac{1}{2} \normsz[\Big]{ \sum_m \mb{h}_m \ast \mb{g} \ast \mb{a}_m - \mb{s} }_W^2 + \nonumber \\
& \lambda_a \sum_m \left(\norm{\mb{a}_m}_1 - \norm{\mb{a}_m}_2\right) +
\nonumber \\
& \frac{\lambda_g}{2} \norm{
 \sqrt{(\mb{c}_0 \ast \mb{g})^2 + (\mb{c}_1  \ast \mb{g})^2}}_2^2 +
\iota_{C}(\mb{g})
\;,
\label{eq:cdlext}
\end{align}
where $\norm{\cdot}_W^2$ denotes an $\ell_2$ norm with weighting in the frequency-domain, $\lambda_a$ and $\lambda_g$ are regularization parameters, $\norm{\mb{a}_m}_1 - \norm{\mb{a}_m}_2$ is the $\ell_1 \!-\! \ell_2$ norm of $\mb{a}_m$, $\mb{c}_0$ and $\mb{c}_1$ are filters that compute the gradients along
image rows and columns respectively, and $\iota_{C}(\cdot)$ is the indicator function\footnote{
The indicator function of set $C$ is defined as
\begin{equation*}
\iota_C(\mb{x}) = \left\{ \begin{array}{ll}
    0 & \text{ if } \mb{x} \in C\\
    \infty & \text{ if } \mb{x} \notin C
    \end{array} \right. \;.
\end{equation*}
} of constraint set $C = \{\mb{x} \in \mbb{R}^N \, | \, \norm{\mb{x}} = 1 \,, x_i \geq 0\,\;\forall i \in \{0, 1, \ldots, N-1 \} \}$.
As is usual for CDL problems, we tackle this bi-convex problem via alternating minimization over the two convex sub-problems corresponding to holding $\mb{g}$ constant and minimizing with respect to the $\mb{a}_m$, and vice-versa. While there has been some work on establishing convergence guarantees for alternating minimization algorithms for dictionary learning~\cite{chatterji-2017-alternating}, we are not aware of any guarantees that would apply to this specific algorithm.

The minimization with respect to the $\mb{a}_m$ can be solved via the ADMM~\cite{boyd-2010-distributed} algorithm for CSC~\cite[Sec. 2.2]{wohlberg-2014-efficient}\cite[Sec. III]{wohlberg-2016-efficient}, with the proximal operator of the $\ell_1$ norm replaced by the proximal operator of the  $\ell_1 - \ell_2$ norm~\cite{lou-2018-fast}, with the required frequency domain weighting being achieved by setting the DC components of the frequency-domain representations of the $\mb{d}_m$ and $\mb{s}$ to zero. The convergence of ADMM applied to problems involving the $\ell_1 - \ell_2$ norm is addressed in~\cite{lou-2018-fast}.

The minimization with respect to $\mb{g}$ can be solved by a variant of the FISTA~\cite{beck-2009-fast} algorithm for the constrained convolutional method of optimal directions (CCMOD)~\cite{garcia-2018-convolutional}. The only changes required to this algorithm are (i) implement the frequency-domain weighting by setting the DC component of the frequency-domain representations of $\sum_m \mb{h}_m \ast \mb{a}_m$ and $\mb{s}$ to zero in the gradient calculation~\cite[Sec. III.D]{garcia-2018-convolutional}, (ii) include a term for the gradient regularization in the calculation of the FISTA gradient, and (iii) compose the usual spatial support projection~\cite[Sec. III.D]{garcia-2018-convolutional} in the FISTA proximal step with a clipping to zero of negative values, which is the proximal operator of the non-negativity constraint. Since this is a convex problem, the usual convergence results for FISTA apply~\cite{beck-2009-fast}.

\section{Results}
\label{sec:rslt}

\subsection{Test Images}
\label{sec:tstdat}

Our benchmark images were simulated to reproduce a realistic distribution of star brightness, pixel sampling, and noise. They span a range of PSF shapes and star densities. The scene consists of PSF light profiles of point sources (stars) on top of a constant sky background. After injecting uniformly distributed stars at random sub-pixel locations and re-sampling to the pixel grid of the image, we add Poisson noise
to model the effects of counting statistics in electro-optical sensors such as CCDs and CMOS arrays. The amplitude of the signal (full 16-bit dynamic range) and sky background (a flat DC offset of 1000 counts) are typical of well exposed astronomical images, where the noise distribution is effectively Gaussian. The baseline noise level corresponds to an inverse gain of 1 electron per data number (variance equal to signal).

The observed number density of stars varies dramatically across the sky. This, in combination with the field of view, sensitivity, and the spatial extent of the PSF, will determine the severity of source confusion.  The density of stars in our test images (see~\fig{testimage} in~\appx{sptstdat}) varies from 100 to 1 pixels per star, i.e. between 655 and 65,500 stars in a tile of $256 \times 256$ pixels. This size is both sufficiently large for a robust PSF estimate and sufficiently small to avoid significant variations of the PSF and the sky background within the tile.

We use a set of four reference PSFs, shown in~\fig{psfref} in~\appx{sptstdat}. The ``narrow'' PSF consists of a circularly symmetric pseudo-Gaussian function\footnote{A pseudo-Gaussian is an inverse of a Taylor expansion of the exponential used by the DoPHOT software~\cite{schechter-1993-dophot}.} with Full Width at Half Maximum (FWHM) of 2 pixels, resulting in a near critical sampling. This represents a very sharp image under excellent viewing conditions. The ``wide'' PSF has the same shape as the narrow one, except for $\rm FWHM = 4$ pixels. This represents poor focus and/or strong atmospheric blurring. The ``elongated'' PSF is an elliptical pseudo-Gaussian at $45$ degrees with the major and minor axis $\rm FWHM = 4$ and $2$ pixels. An elongated PSF may arise e.g. due to a coma in imaging optics or imperfect tracking of the sidereal sky motion. Finally, the ``complex'' PSF includes one of each with different amplitudes and small centroid offsets to simulate shapes resulting from a combination of factors.

\subsection{Metrics}
\label{sec:metric}

The metric for evaluating the accuracy of sampled estimates of a continuous PSF must take into account both a scaling ambiguity (multiplication of the PSF by a scalar factor can be compensated by dividing the star field by the same factor) and a phase shift ambiguity (a phase shift in the PSF can be compensated by a corresponding phase shift in the star field). We denote the reference continuous PSF by the function $g(\cdot)$ and the sampled PSF with which it is to be compared by vector $\mb{h}$, with components $h_i$, which are assumed to represent samples of an underlying continuous function $h(\cdot)$ taken at points $\mc{I} \subset \mbb{Z}^+$. A correlation function between continuous function $g(\cdot)$ and vector $\mb{h}$ at sampling offset $n$ is defined as
\begin{equation}
c(n) = \frac{\sum_{i \in \mc{I}} h_i g(i + n / N_{\mathrm{R}})}
            {\sqrt{\sum_{i \in \mc{I}} h_i^2}
             \sqrt{\sum_{i \in \mc{I}} g(i + n / N_{\mathrm{R}})^2}} \;,
\end{equation}
where $N_{\mathrm{R}}$ is the sub-pixel resolution factor at which the correlation is computed. Now, defining
\[
\hat{n} = \argmax c(n) \quad \mb{g} = g(\mc{I} + \hat{n} / N_{\mathrm{R}})
 \quad a = \mb{h}^T \mb{h} / \mb{g}^T \mb{h} \;,
\]
we compute the value of the metric as the Signal-to-Noise Ratio (SNR) of $\mb{h}$ with respect to $a \mb{g}$, \ie, the SNR between a sampled and scaled representation of $g(\cdot)$ with the sampling offset, $n$, and scaling, $a$, chosen to maximize the SNR.

\begin{table}
\centering
\caption{PSF estimation performance in SNR (dB) for the RCA method with parameters optimized for each case. Performance relative to that of the proposed method in~\tbl{cdlrule} is indicated by the font and parentheses\protect\footnotemark[6].
}
\label{tbl:rcaopt}
\begin{tabular}{|l|c|c|c|c|c|c|} \hline
\diagbox{shape}{density} & 1 & 10 & 25 & 50 & 100 \\ \hline
narrow    & \lt{20.17} & \lt{23.63} & \lt{20.78} & \lt{20.56} & \lt{23.77} \\ \hline
wide      & \lt{24.39} & \lt{25.91} & \lt{25.43} & \lt{24.57} & \gt{26.42} \\ \hline
elongated     & \lt{23.18} & \lt{26.11} & \lt{22.76} & \lt{22.33} & \lt{23.75} \\ \hline
complex   & \lt{28.45} & \lt{26.12} & \lt{25.08} & \lt{24.77} & \gt{25.38} \\ \hline
\end{tabular}
\end{table}

\footnotetext[6]{Performance relative to values in the other table is indicated by parentheses  where the performance is less than 2dB better than that in the other table, and by \textbf{bold font} where it is at least 2dB better than that in the other table.}

\begin{table}
\centering
\caption{
PSF estimation performance in SNR (dB) for the proposed method. Performance relative to that of the RCA method in~\tbl{rcaopt} is indicated by the font and parentheses\protect\footnotemark[6].
}
\label{tbl:cdlrule}
\begin{tabular}{|l|c|c|c|c|c|c|} \hline
\diagbox{shape}{density} & 1 & 10 & 25 & 50 & 100 \\ \hline
narrow   &   \mgt{34.39} & \mgt{39.06} & \mgt{36.15} & \mgt{36.36} & \mgt{31.57} \\ \hline
wide     &   \mgt{34.41} & \mgt{32.97} & \mgt{30.90} & \mgt{31.19} & \lt{25.46} \\ \hline
elongated    &   \mgt{33.20} & \mgt{35.14} & \mgt{34.11} & \mgt{34.95} & \mgt{30.95} \\ \hline
complex  &   \mgt{30.32} & \mgt{29.52} & \mgt{27.24} & \mgt{29.71} & \lt{25.08} \\ \hline
\end{tabular}
\end{table}

\subsection{Performance Comparisons}
\label{sec:perfcmp}

A direct comparison to existing	approaches is difficult	because	few algorithms can handle extreme crowding and even fewer have publicly available implementations. We compare the performance of the proposed algorithm with that of the recent Resolved Components Analysis (RCA)~\cite{ngole-2016-constraint} method, using the implementation provided by the authors~\cite{ngole-2017-rca}. The algorithm takes input in the form of postage stamp images approximately centered around well detected, isolated stars. In our most crowded images, finding isolated stars is virtually impossible. In order to ensure the best possible quality of input data, we manually selected several dozen bright stars, while attempting to minimize the contamination from neighboring objects. Since this method has six parameters for which there are no clear selection guidelines, for each test case we select the best parameters by evaluating the performance of the method over 9000 different parameter combinations. The results of this experiment are displayed in~\tbl{rcaopt}.

The proposed algorithm is implemented in Python as an extension~\cite{wohlberg-2021-cdlpsf} of the SPORCO package~\cite{wohlberg-2016-sporco, wohlberg-2017-sporco}. Parameter $M$ was set to 5 for all cases, $K$ and $\sigma_0$ were chosen according to the PSF shape, and the remaining parameters were chosen according to the star density, as described in~\appx{param}.\footnote{The development of reliable automated parameter selection, which would enhance the practical value of the proposed method, is left as a topic for future study.} The results of this experiment are displayed in~\tbl{cdlrule}. Despite the much larger parameter space explored in computing the RCA results, the performance of the proposed method exceeds that of RCA by more than 2db for all but two cases, and in some cases is better by more than 10 dB. The only cases where RCA outperforms the proposed method are at the lowest star density of 100 pixels per star.

\section{Conclusions}
\label{sec:cncl}

We have proposed a new PSF estimation algorithm, based on a CDL framework, for crowded astronomical imagery.  The resulting performance over a very wide range of crowding conditions compares very favorably with that of RCA, a recent alternative method. Unlike competing algorithms, our approach does not require laborious pre-processing to select isolated stars. The need to detect and model individual point sources---a complicated and error prone task---is eliminated altogether. Our hypothesis is that the global nature of the proposed model accounts for most of the observed performance improvements over the usual patch-based methods. The CDL method can be further extended to support a spatial mask for rejection of artifacts such as saturated pixels, cosmic ray hits, or bad columns. These properties make the algorithm well suited for PSF estimation anywhere from extremely crowded stellar populations like the Galactic bulge and globular clusters to more routine work.

\bibliographystyle{IEEEtranD}
\bibliography{psfest}

\appendices

\newpage

\section{CDL Algorithm}
\label{sec:cdlalg}

The algorithm for minimization of our CDL problem,~\eq{cdlext}, consists of alternating minimization with respect to the $\mb{a}_m$ (sparse coding) and to $\mb{g}$ (dictionary update).

\subsection{Sparse Coding}
\label{sec:cdlsc}

The minimization with respect to the $\mb{a}_m$ can be expressed as
\begin{align}
 \argmin_{\{\mb{a}_m\}} \; & \frac{1}{2} \normsz[\Big]{ \sum_m \mb{d}_m \ast \mb{a}_m - \mb{s} }_W^2 + \sum_m \iota_C(\mb{a}_m)  \, + \nonumber \\
& \lambda_a \sum_m \left(\norm{\mb{a}_m}_1 - \norm{\mb{a}_m}_2\right)
\;,
\label{eq:csc}
\end{align}
where $\mb{d}_m = \mb{h}_m \ast \mb{g}$. This problem is similar to the standard convolutional sparse coding (CSC)~\cite{wohlberg-2016-efficient} problem, and can be solved via a variant of the ADMM algorithm described in~\cite{wohlberg-2016-efficient}
\begin{align}
  \{\mb{a}_m\}^{(j+1)} &= \argmin_{\{\mb{a}_m\}} \frac{1}{2}
  \normsz[\Big]{\sum_m \mb{d}_m \ast \mb{a}_m - \mb{s}}_W^2 + \nonumber \\ &
  \hspace{5em} \frac{\rho_a}{2} \sum_m \norm{
    \mb{a}_m - \mb{u}_m^{(j)} + \mb{v}_m^{(j)}}_2^2 \label{eq:bpdnxprob} \\
  \{\mb{u}_m\}^{(j+1)} &= \argmin_{\{\mb{u}_m\}} \lambda_a \sum_m
  \left(\norm{\mb{u}_m}_1 - \norm{\mb{u}_m}_2\right) + \nonumber \\ & \hspace{5em} \sum_m \iota_{C_a}(\mb{u}_m)  +\nonumber \\ & \hspace{5em} \frac{\rho_a}{2}
  \sum_m \norm{ \mb{a}_m^{(j+1)} - \mb{u}_m +
    \mb{v}_m^{(j)}}_2^2 \label{eq:bpdnyprob}  \\
  \mb{v}_m^{(j+1)} &= \mb{v}_m^{(j)} + \mb{a}_m^{(j+1)} -
  \mb{u}_m^{(j+1)} \; , \label{eq:bpdnuprob}
\end{align}
where $\rho_a$ is the ADMM penalty parameter that controls the convergence of the algorithm.

Update~\eq{bpdnxprob} can be solved by setting the DC components of the frequency-domain representations of the $\mb{d}_m$ and $\mb{s}$ to zero before applying the computationally efficient frequency-domain solution described in~\cite[Sec. 2.2]{wohlberg-2014-efficient}\cite[Sec. III]{wohlberg-2016-efficient}. Update~\eq{bpdnyprob} corresponds to the proximal operators of the $\ell_1 - \ell_2$ norm, for which there is a closed form expression~\cite{lou-2018-fast}.

\subsection{Dictionary Update}
\label{sec:cdldu}

The minimization with respect to $\mb{g}$ can be expressed as
\begin{align}
 \argmin_{\mb{g}} \; & \frac{1}{2} \normsz[\Big]{\mb{b} \ast \mb{g} - \mb{s}}_W^2 +  \iota_C(\mb{g}) + \nonumber \\
& \frac{\lambda_g}{2} \norm{
 \sqrt{(\mb{c}_0 \ast \mb{g})^2 + (\mb{c}_1  \ast \mb{g})^2}}_2^2
\;,
\label{eq:ccmod}
\end{align}
where $\mb{b} = \sum_m \mb{h}_m \ast \mb{a}_m$,
which is a regularized and constrained  deconvolution of $\mb{s}$ with respect to $\mb{b}$. This problem is similar to the constrained convolutional method of optimal directions (CCMOD)~\cite{garcia-2018-convolutional} problem, and can be solved via a variant of the FISTA algorithm described in~\cite[Sec. III.D]{garcia-2018-convolutional}
\begin{align}
  \mb{g}^{(i+1)} &= \prox_{\iota_{C}}\bigg( \mb{y}^{(i)} - L_g^{-1} \nabla_{\mb{y}} f(\mb{y}) \bigg) \label{eq:ccmodyfista}  \\
  t^{(i+1)} & = \frac{1}{2} \bigg(1 + \sqrt{1 + 4 \, (t^{(i)})^2} \bigg) \label{eq:ccmodtfista} \\
  \mb{y}^{(i+1)} &= \mb{g}^{(i+1)} +  \frac{t^{(i)} - 1}{t^{(i+1)}} \Big(\mb{g}^{(i+1)} - \mb{y}^{(i)} \Big) \;, \label{eq:ccmoddfista}
\end{align}
where $f(\mb{g})$ represents the sum of the first and third terms in~\eq{ccmod},
$t^{(0)} = 1$, and $L_g > 0$ is a parameter controlling the step size. The frequency-domain weighting of the data fidelity term can be implemented by setting the DC component of the frequency-domain representations of $\sum_m \mb{h}_m \ast \mb{a}_m$ and $\mb{s}$ to zero in the calculation of the gradient of $f(\mb{g})$, and the proximal operator of the indicator function of $C$ corresponds to the composition of the usual spatial support projection~\cite[Sec. III.D]{garcia-2018-convolutional} in the FISTA proximal step with clipping to zero of negative values and normalization.

\subsection{Alternating Minimization}
\label{sec:cdlam}

\begin{algorithm}[htpb]
 \dontprintsemicolon
 \SetKw{init}{Initialize:}
 \SetKw{pre}{Precompute:}
 \KwIn{image $\mb{s}$ \;}
 \init Initialize $\mb{g}$ as a symmetric Gaussian PSF of width $\sigma_0$ \;
 \For{$i \in {1, 2, \ldots, N_{\text{iter},0}}$}{
   Compute sparse coding steps~\eq{bpdnxprob}--\eq{bpdnuprob} with fixed dictionary $\mb{d}_m = \mb{h}_m \ast \mb{g}$ \;
  Set final $\mb{a}_m$ as the current sparse representation \;
  }
  \For{$i \in {1, 2, \ldots, N_{\text{iter},0}}$}{
   Compute dictionary update steps~\eq{ccmodyfista}--\eq{ccmoddfista} with fixed $\mb{b} = \sum_m \mb{h}_m \ast \mb{a}_m$  \;
   Set final $\mb{g}$ as the current PSF estimate \;
  }
  \For{$i \in {1, 2, \ldots, N_{\text{iter}}}$}{
   Compute sparse coding steps~\eq{bpdnxprob}--\eq{bpdnuprob} with fixed dictionary $\mb{d}_m = \mb{h}_m \ast \mb{g}$ \;
   Set resulting $\mb{a}_m$ as the current sparse representation \;
   Compute dictionary update steps~\eq{ccmodyfista}--\eq{ccmoddfista} with fixed $\mb{b} = \sum_m \mb{h}_m \ast \mb{a}_m$ \;
   Set resulting $\mb{g}$ as the current PSF estimate \;
  }
 \KwOut{Estimated PSF $\mb{g}$ \;}
 \vspace{2mm}
 \caption{Summary of CDL algorithm for PSF estimation.}
 \label{alg:cdl}
\end{algorithm}

The full CDL algorithm is summarized in Alg.~\ref{alg:cdl}.

\section{Parameter Selection}
\label{sec:param}

\begin{figure}[htpb]
  \includegraphics[width=0.48\textwidth]{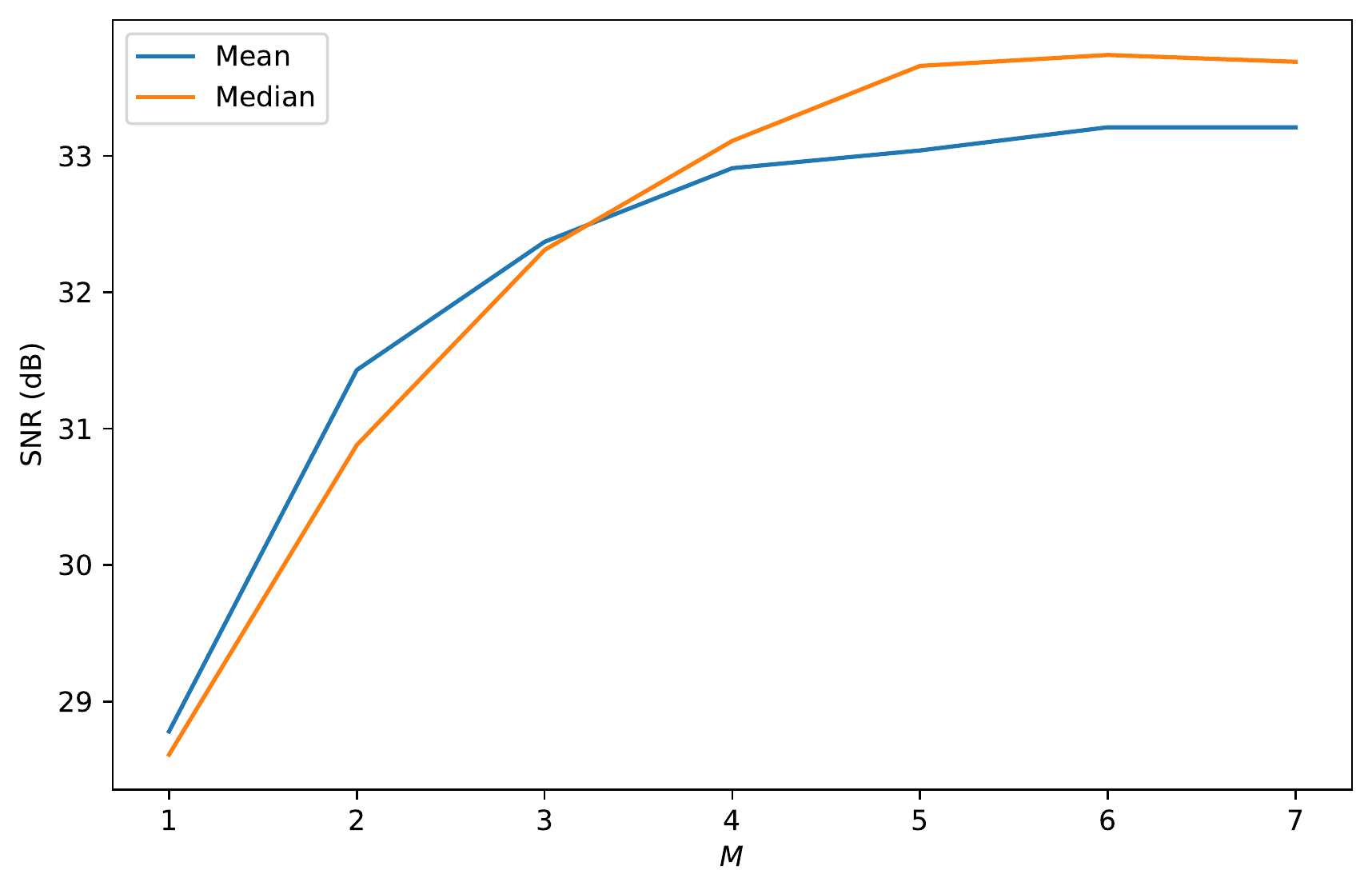}
  \caption{Dependence on parameter $M$ of mean and median of PSF estimation performance over all test cases, with parameters $\sigma_0$, $\lambda_a$, $\lambda_g$, $\rho_a$, and $L_g$ individually optimized for each case. Note that competitive performance is even achieved at $K = 1$, which corresponds to a dictionary with a single filter, without any interpolation to account for sub-pixel offsets of the PSF.}
  \label{fig:snrm}
\end{figure}

Our algorithm has four model parameters $M$ (number of sub-pixel offsets of the fundamental PSF $\mb{g}$), $K$ (order of the Lanczos interpolation used in computing the sub-pixel shifts), $\lambda_a$ (regularization parameter for the sparse representation), and $\lambda_g$ (regularization parameter for the fundamental PSF $\mb{g}$). In addition, there are five optimization parameters $\sigma_0$ (width parameter of the symmetric Gaussian PSF used to initialize the dictionary learning), $\rho_a$ (penalty parameter of the ADMM algorithm of the CSC update), $L_g$ (inverse step length parameter of the FISTA algorithm for the dictionary update), $N_{\text{iter},0}$ (initialization iterations), and $N_{\text{iter}}$ (main iterations). We set $M = 5$ for all our experiments since this value was found to give represent a good balance between performance (see~\fig{snrm}) and computational cost, which is quadratic in $M$ for 2D signals. We set $K = 5$ for ``complex'' and ``narrow'' PSF shapes, and $K = 10$ for the ``elong'' and ``wide'' shapes since these values maximize the accuracy of the Lanczos kernel in interpolating the respective PSF shapes.

For the results in~\tbl{cdlrule}, we set $\sigma_0 = 1.0$ for the ``complex'' and ``wide'' PSF shapes, $\sigma_0 = 0.5$ for the ``narrow'' and ``elong'' shapes, $N_{\text{iter},0} = 10$, and $N_{\text{iter}} = 100$. The other parameters are all selected according to the star density, as indicated in~\tbl{cdlparamrule}. The dependency of $\sigma_0$ on the PSF shape and of the other parameters on the star density was chosen by selecting the dependency rules to maximize the mean SNR for all test cases over a set of 768 different parameter combinations.

\begin{table}[htb]
\centering
\caption{Parameter selection according to star density.}
\label{tbl:cdlparamrule}
\begin{tabular}{|l|c|c|c|c|c|c|} \hline
\diagbox{param.}{density} & 1 & 10 & 25 & 50 & 100 \\ \hline
$\lambda_a$  &   0.01 & 0.01 & 0.01 & 0.01 & 0.1 \\ \hline
$\lambda_g$  &   0.01 & 0.1 & 0.1 & 0.1 & 0.1 \\ \hline
$\rho_a$    &    1 &  1&  1 &  1 & 10 \\ \hline
$L_g$      &     50 &  100 &  100 &  500 & 1000 \\ \hline
\end{tabular}
\end{table}

\section{Reference PSFs and Test Images}
\label{sec:sptstdat}

\begin{figure}[htpb]
  \begin{subfigure}{.24\textwidth}
  \centering
  \includegraphics[width=\textwidth]{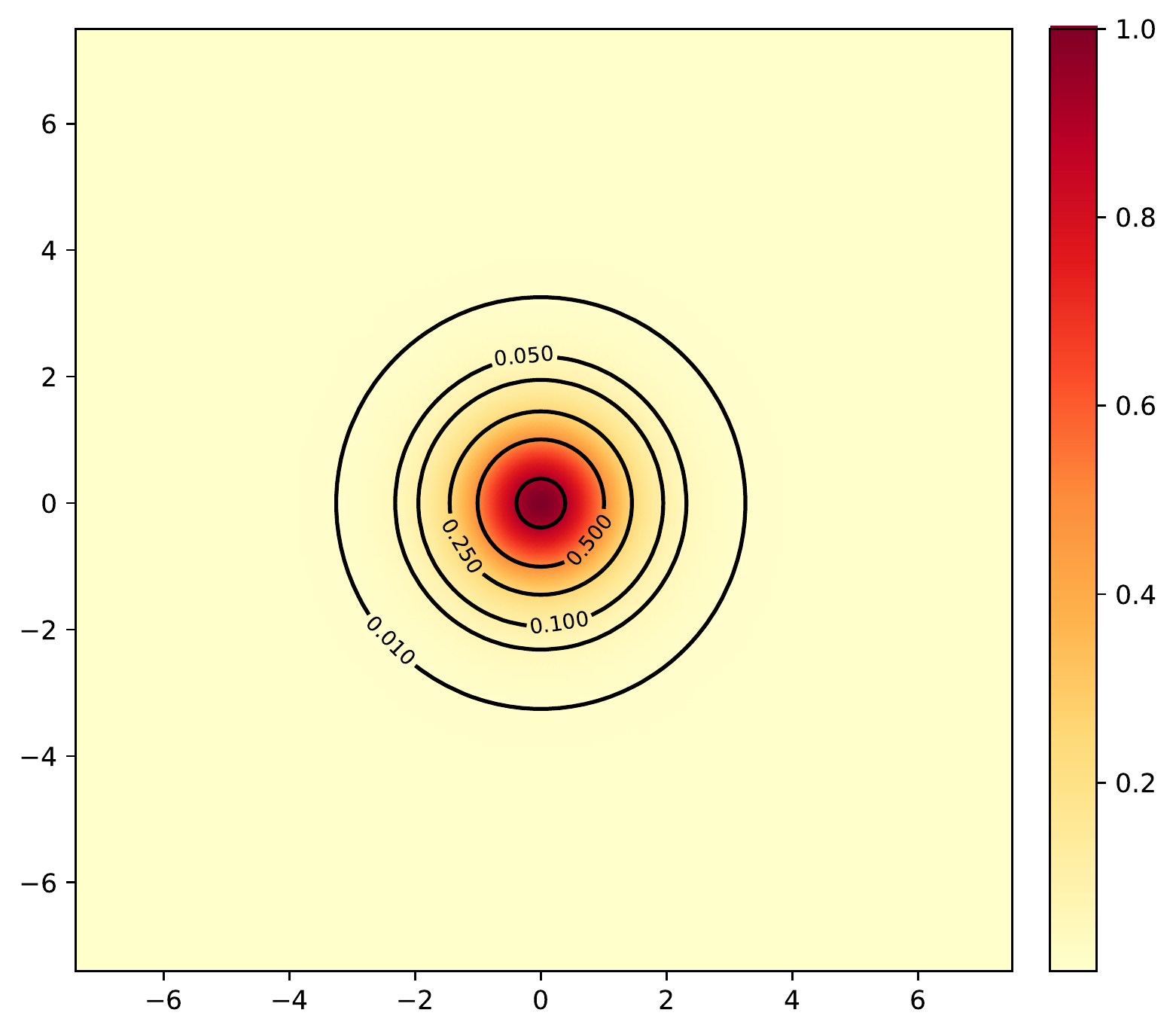}
  \caption{Narrow PSF}
  \label{fig:psfnrw}
  \end{subfigure}
  \begin{subfigure}{.24\textwidth}
  \centering
  \includegraphics[width=\textwidth]{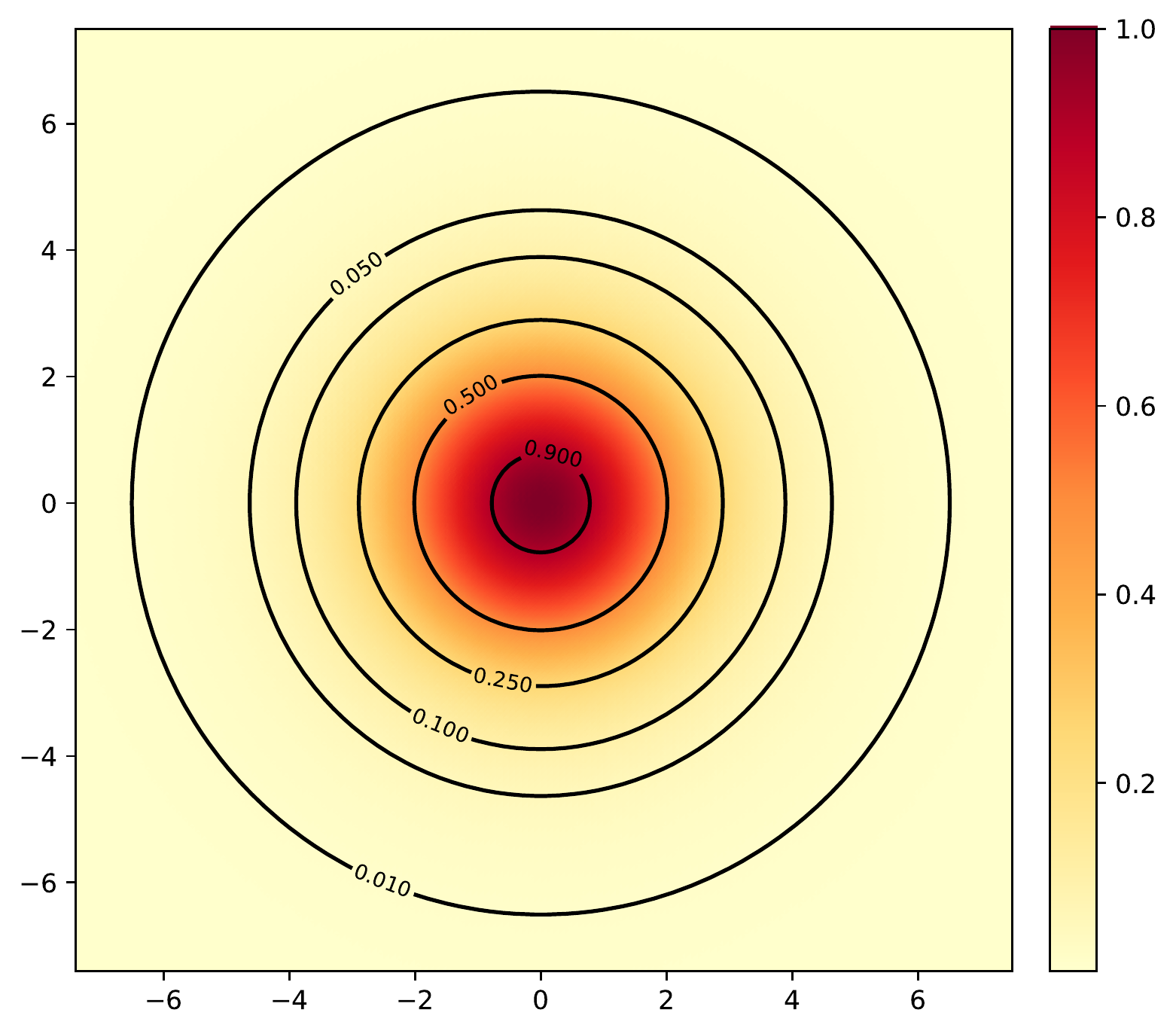}
  \caption{Wide PSF}
  \label{fig:psfwide}
  \end{subfigure} \\
  \begin{subfigure}{.24\textwidth}
  \centering
  \includegraphics[width=\textwidth]{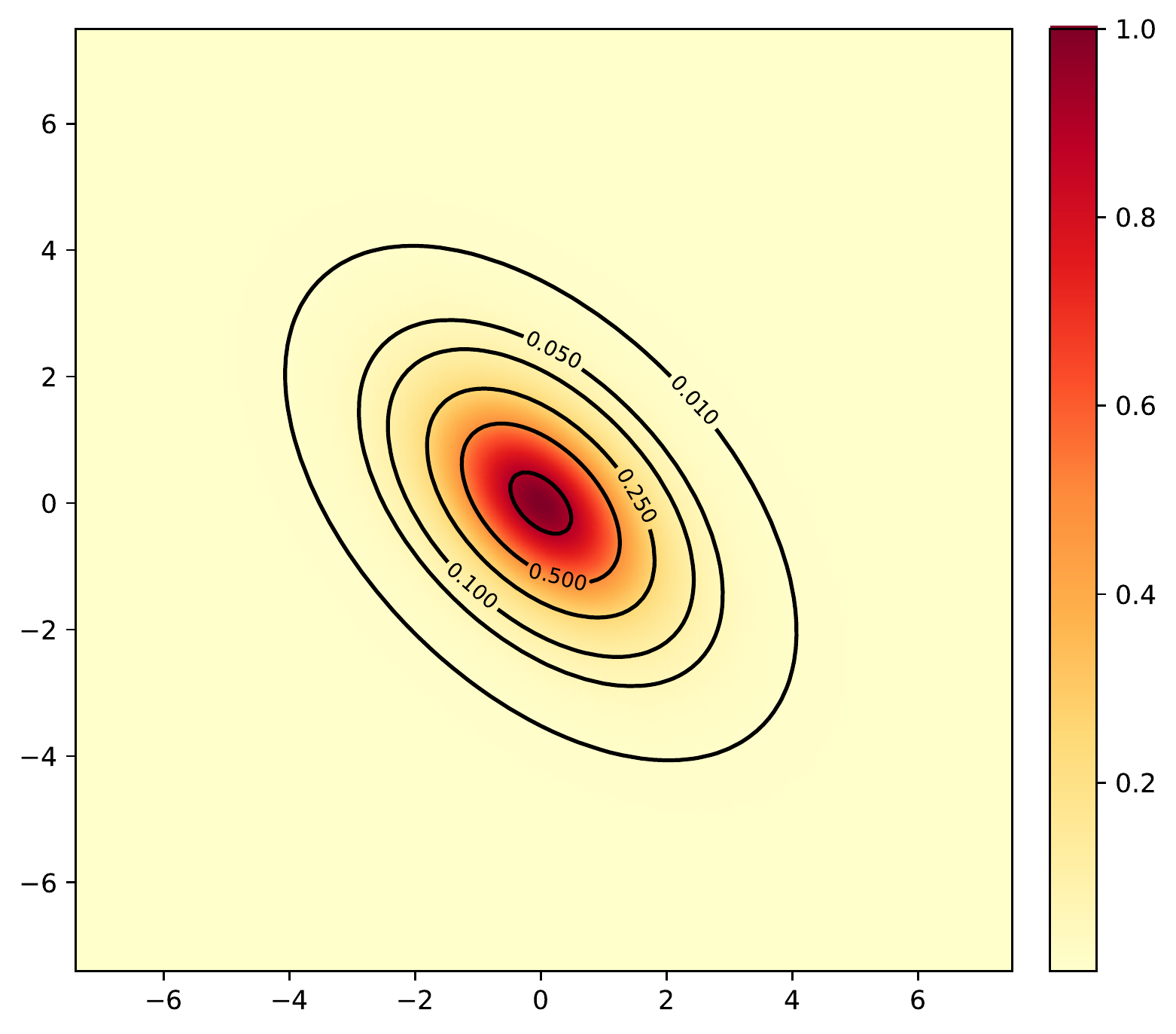}
  \caption{Elongated PSF}
  \label{fig:psfelng}
  \end{subfigure}
  \begin{subfigure}{.24\textwidth}
  \centering
  \includegraphics[width=\textwidth]{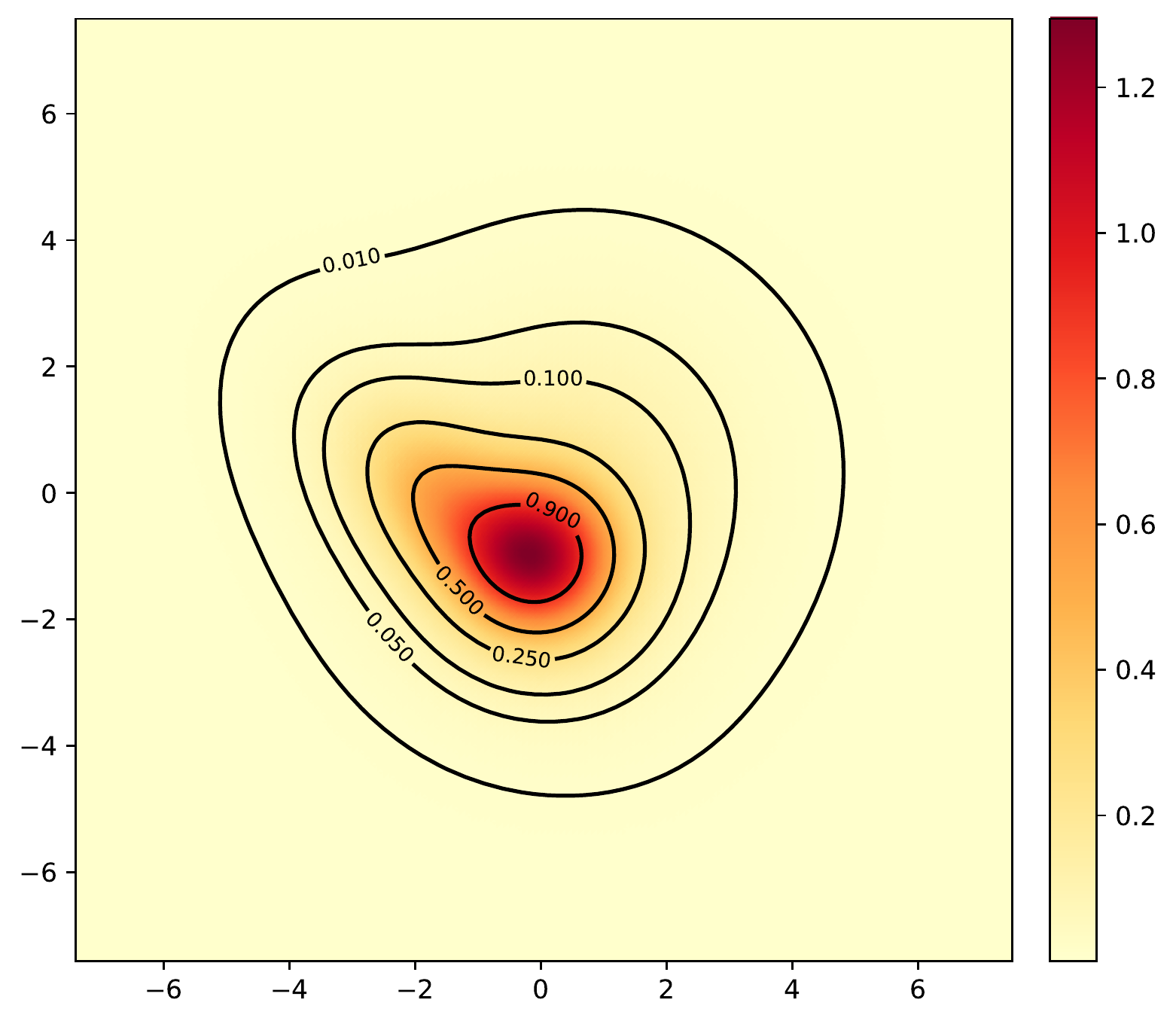}
  \caption{Complex PSF}
  \label{fig:psfcmplx}
  \end{subfigure}
  \caption{Reference PSFs}
  \label{fig:psfref}
  \vspace{-3mm}
\end{figure}

\begin{figure}[htb]
  \begin{subfigure}{0.24\textwidth}
  \centering
  \includegraphics[width=0.9\textwidth]{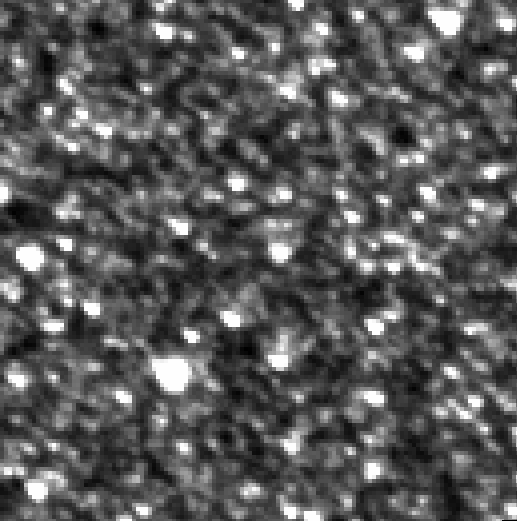}
  \caption{1 pixel per star}
  \label{fig:testimage_d1}
  \end{subfigure}
  \begin{subfigure}{0.24\textwidth}
  \centering
  \includegraphics[width=0.9\textwidth]{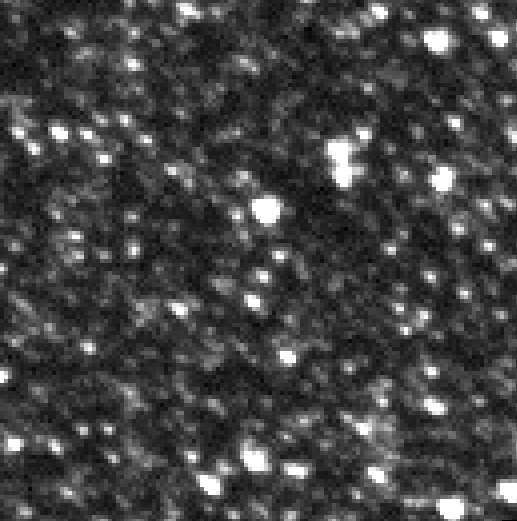}
  \caption{10 pixels per star}
  \label{fig:testimage_d10}
  \end{subfigure}
  \begin{subfigure}{0.24\textwidth}
  \centering
  \includegraphics[width=0.9\textwidth]{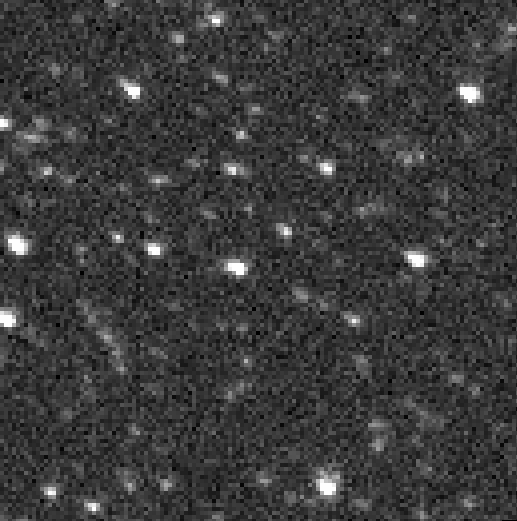}
  \caption{100 pixels per star}
  \label{fig:testimage_d100}
  \end{subfigure}
  \caption{Example test images\protect\footnotemark \hspace{1pt} with noise level 1.0.}
  \label{fig:testimage}
\end{figure}

\footnotetext[7]{The gray scale levels have been adjusted using the ``zscale'' algorithm, which preferentially displays data around the peak of the pixel distribution without computing the actual histogram (see the documentation for the IRAF display utility at \url{https://iraf.net/irafhelp.php?val=display&help=Help+Page\#s\_zscale_algorithm}), and is widely used in the astronomical community.}

Reference PSFs and test images are shown in~\figs{psfref} and~\fign{testimage} respectively.

\section{Computational Cost Comparison}
\label{sec:runtime}

\begin{figure}[htpb]
  \includegraphics[width=0.48\textwidth]{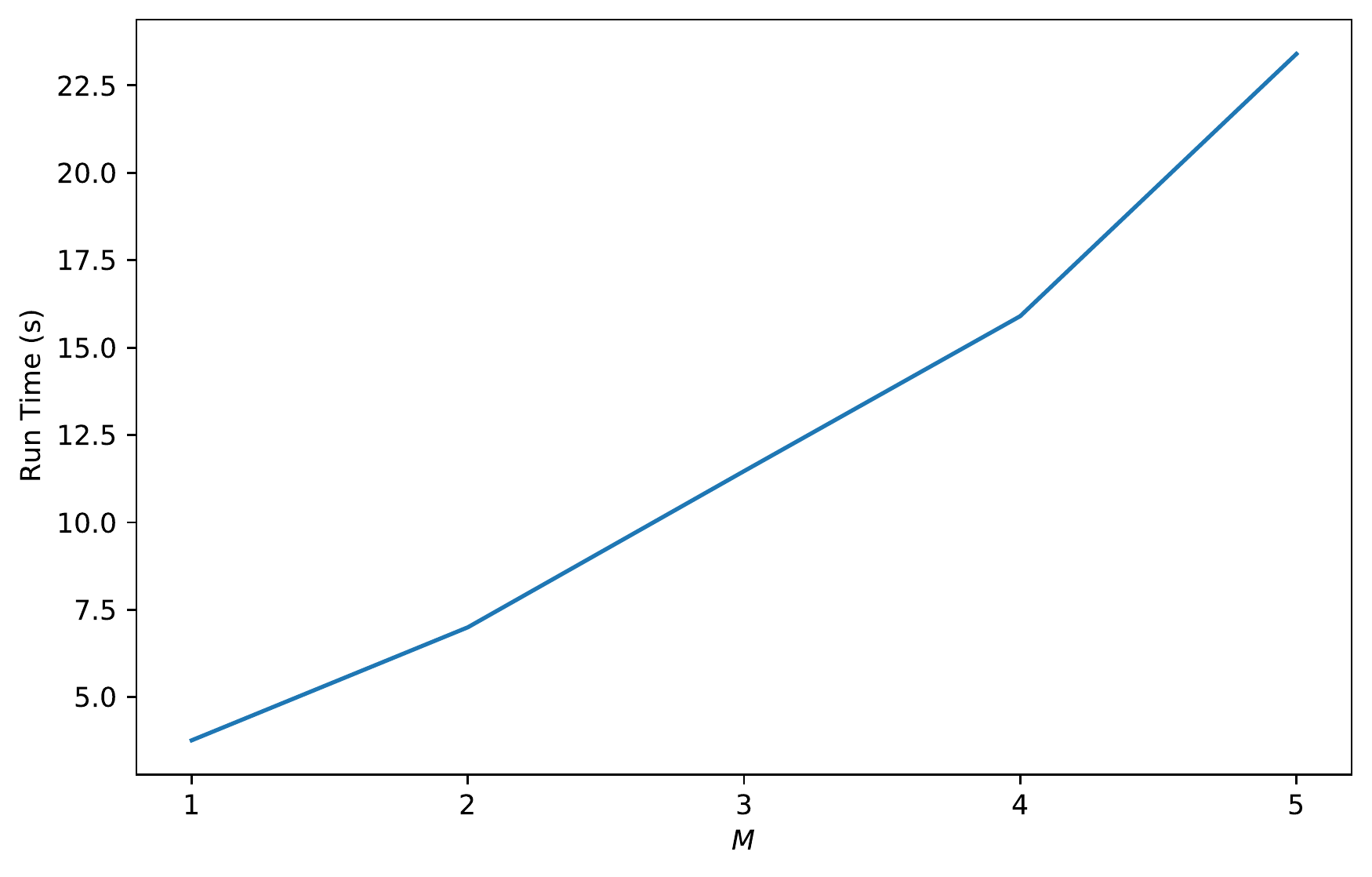}
  \caption{Dependence on parameter $M$ of run time of the proposed method.}
  \label{fig:timem}
\end{figure}

We compared the run times of the proposed method and RCA on a host with a 14 core Xeon E5-2690 CPU. The typical run time of RCA was approximately 6s, and the typical run times of the proposed method, which depend on parameter $M$, are displayed in~\fig{timem}. While the typical run time of the proposed method is approximately 24s when $M=5$, which is the value selected for the results reported in~\tbl{cdlrule} in the main document, smaller values of $M$ have corresponding run times that are closer to that of RCA while retaining good PSF estimation performance (see~\fig{snrm}). It is also important to note that the typical run-time reported for RCA excludes the time required for identifying isolated stars and extracting their surrounding patches, which can be a time-consuming manual process, while no such process is required by the proposed method.

\section{Performance Comparisons}
\label{sec:supperfcmp}

\begin{figure*}[htpb]
  \centering
  \includegraphics[width=\textwidth]{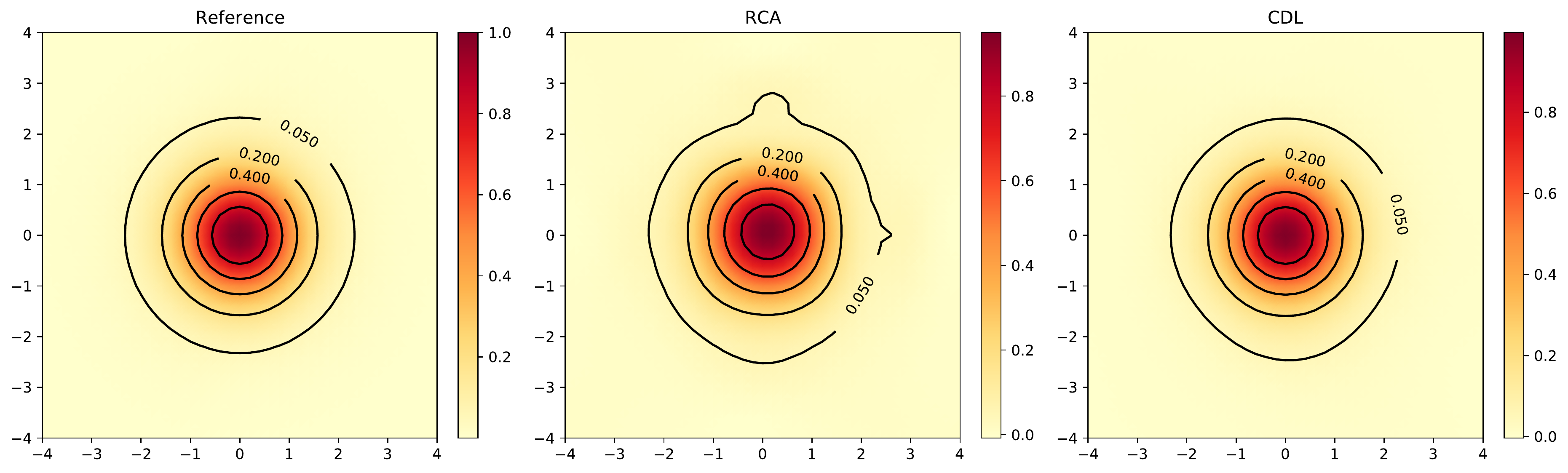}
  \caption{Contour plots comparing the reference ``narrow'' shape PSF with estimates computed via RCA and CDL from images with a star density of 1 pixel per star.}
  \label{fig:psf_narrow_cntr}
\end{figure*}

\begin{figure*}[htpb]
  \centering
  \includegraphics[width=0.95\textwidth]{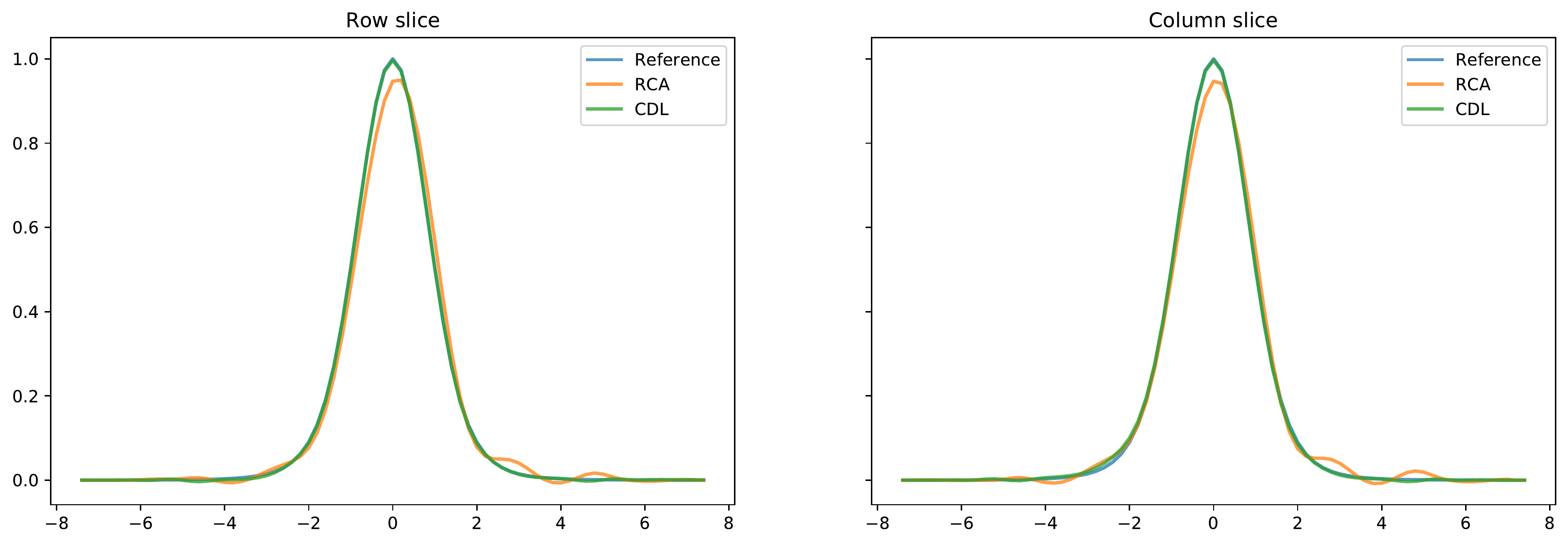}
  \caption{Row and column slices comparing the reference ``narrow'' shape PSF with estimates computed via RCA and CDL from images with a star density of 1 pixel per star.}
  \label{fig:psf_narrow_sect}
\end{figure*}

\begin{figure*}[htpb]
  \centering
  \includegraphics[width=0.95\textwidth]{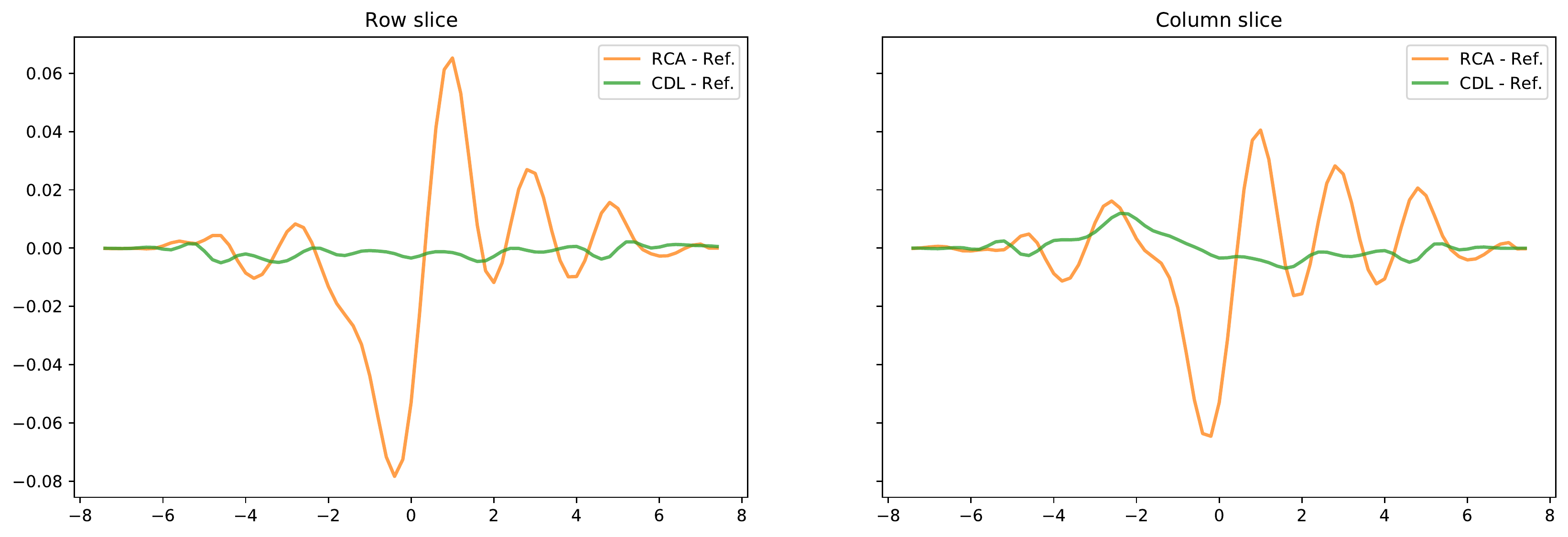}
  \caption{Row and column slices of the differences between the reference ``narrow'' shape PSF and the estimates (a constant zero difference represents a perfect estimate) computed via RCA and CDL from images with a star density of 1 pixel per star.}
  \label{fig:psf_narrow_secdif}
\end{figure*}

\begin{figure*}[htpb]
  \centering
  \includegraphics[width=\textwidth]{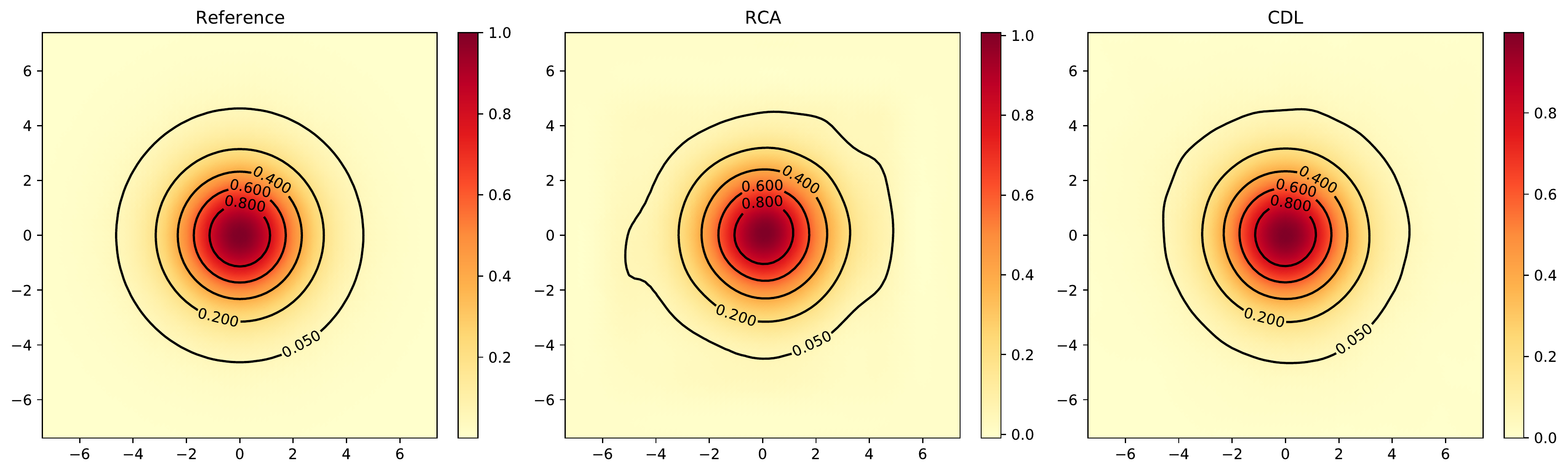}
  \caption{Contour plots comparing the reference ``wide'' shape PSF with estimates computed via RCA and CDL from images with a star density of 1 pixel per star.}
  \label{fig:psf_wide_cntr}
\end{figure*}

\begin{figure*}[htpb]
  \centering
  \includegraphics[width=0.95\textwidth]{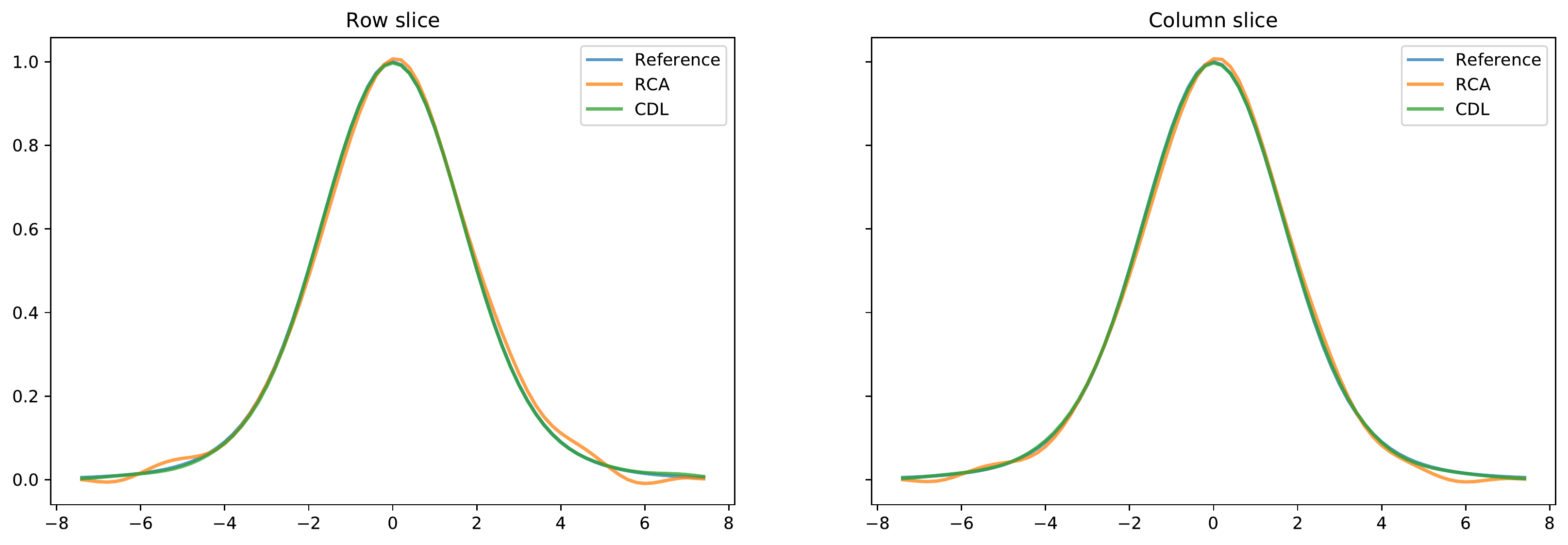}
  \caption{Row and column slices comparing the reference ``wide'' shape PSF with estimates computed via RCA and CDL from images with a star density of 1 pixel per star.}
  \label{fig:psf_wide_sect}
\end{figure*}

\begin{figure*}[htpb]
  \centering
  \includegraphics[width=0.95\textwidth]{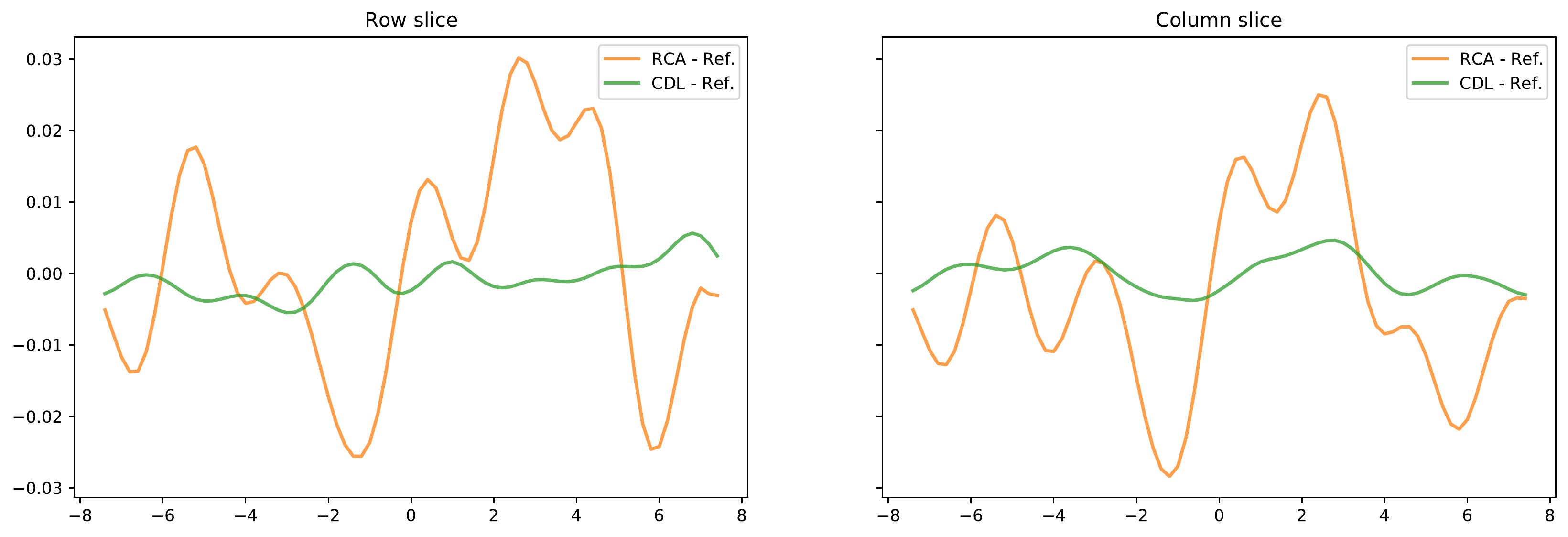}
  \caption{Row and column slices of the differences between the reference ``wide'' shape PSF and the estimates (a constant zero difference represents a perfect estimate) computed via RCA and CDL from images with a star density of 1 pixel per star.}
  \label{fig:psf_wide_secdif}
\end{figure*}

\begin{figure*}[htpb]
  \centering
  \includegraphics[width=\textwidth]{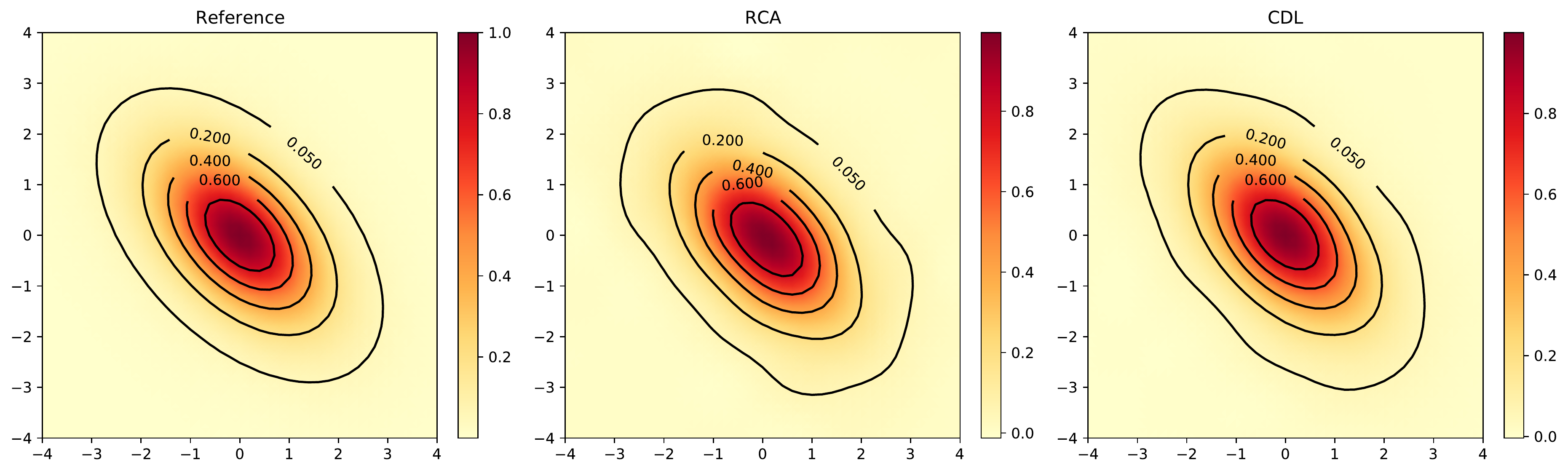}
  \caption{Contour plots comparing the reference ``elong'' shape PSF with estimates computed via RCA and CDL from images with a star density of 1 pixel per star.}
  \label{fig:psf_elong_cntr}
\end{figure*}

\begin{figure*}[htpb]
  \centering
  \includegraphics[width=0.95\textwidth]{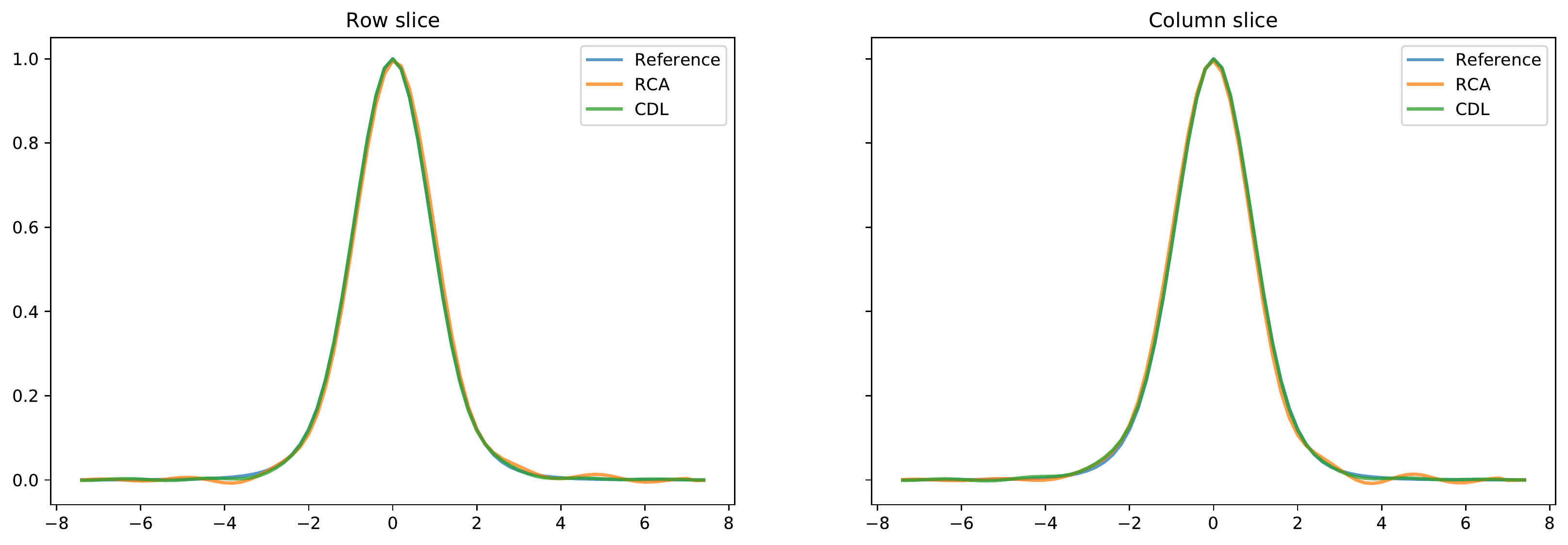}
  \caption{Row and column slices comparing the reference ``elong'' shape PSF with estimates computed via RCA and CDL from images with a star density of 1 pixel per star.}
  \label{fig:psf_elong_sect}
\end{figure*}

\begin{figure*}[htpb]
  \centering
  \includegraphics[width=0.95\textwidth]{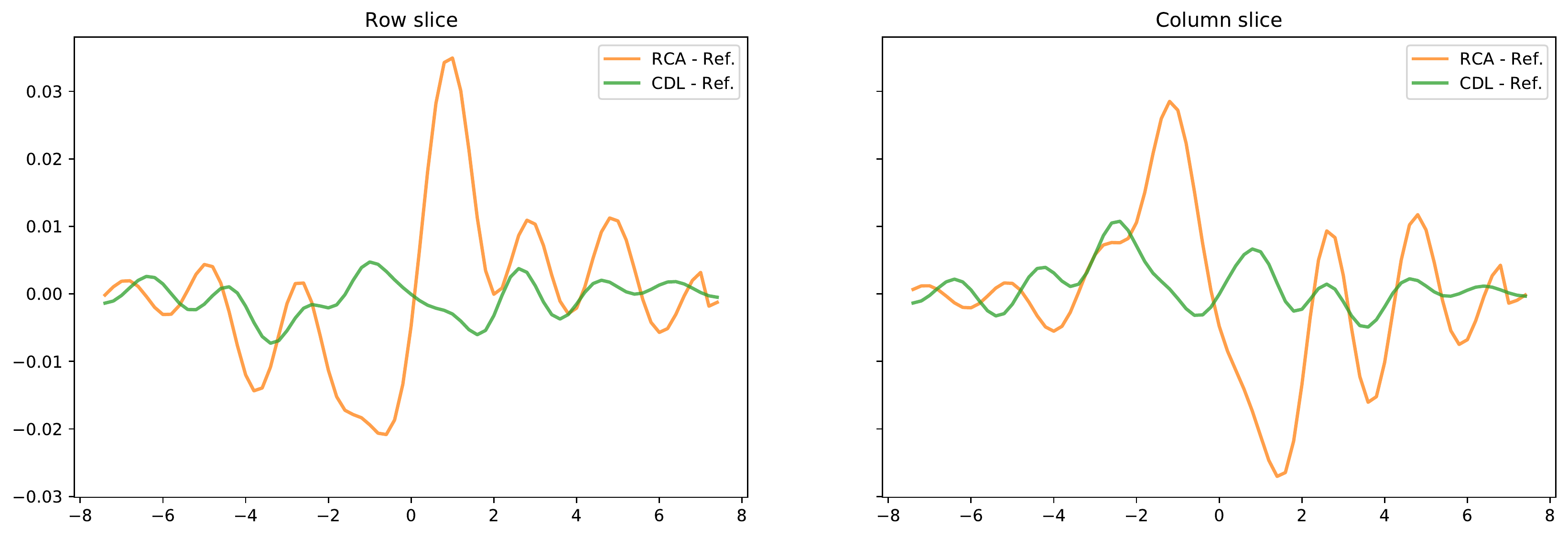}
  \caption{Row and column slices of the differences between the reference ``elong'' shape PSF and the estimates (a constant zero difference represents a perfect estimate) computed via RCA and CDL from images with a star density of 1 pixel per star.}
  \label{fig:psf_elong_secdif}
\end{figure*}

\begin{figure*}[htpb]
  \centering
  \includegraphics[width=\textwidth]{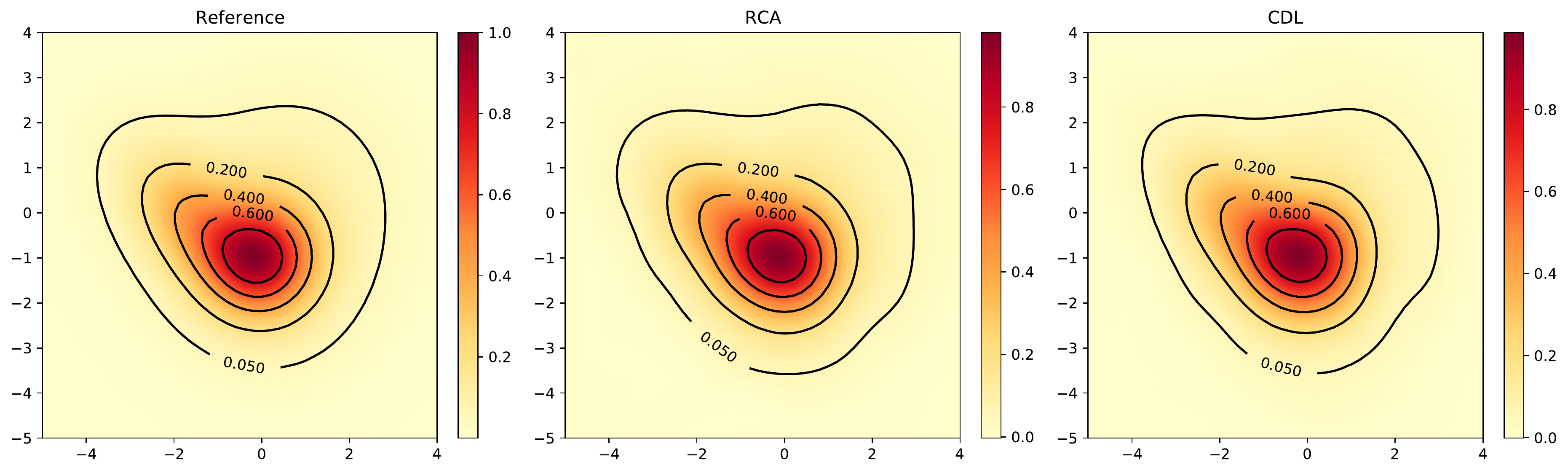}
  \caption{Contour plots comparing the reference ``complex'' shape PSF with estimates computed via RCA and CDL from images with a star density of 1 pixel per star.}
  \label{fig:psf_complex_cntr}
\end{figure*}

\begin{figure*}[htpb]
  \centering
  \includegraphics[width=0.95\textwidth]{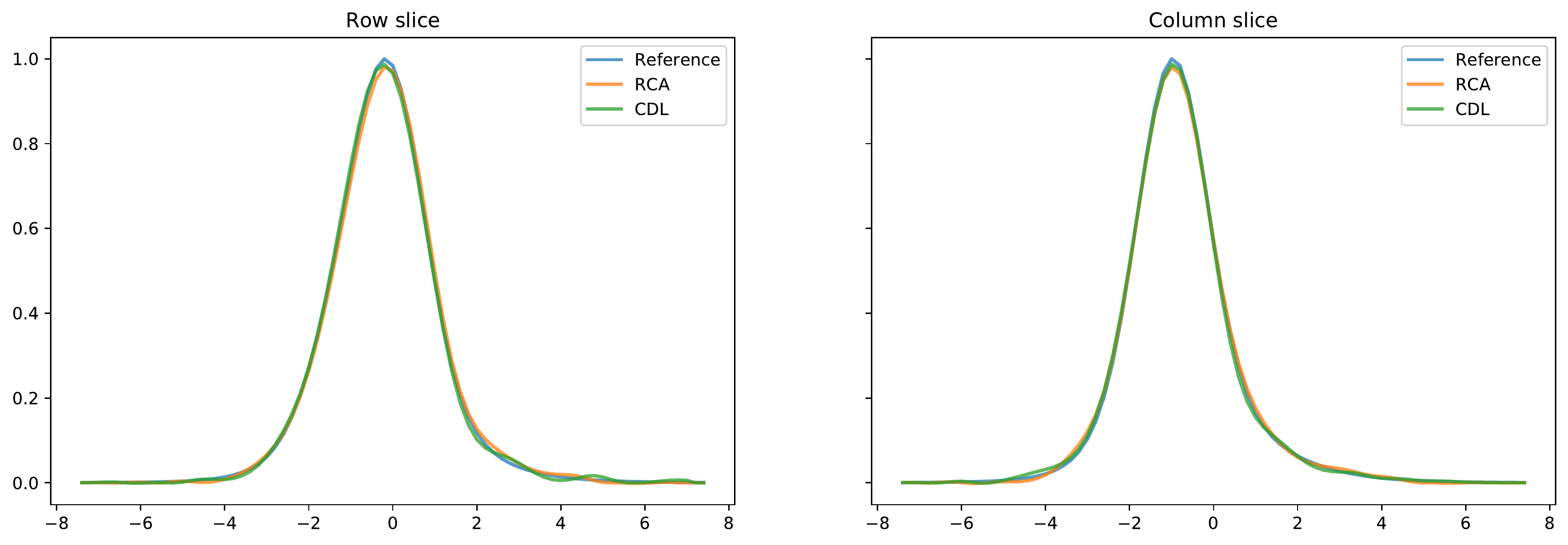}
  \caption{Row and column slices comparing the reference ``complex'' shape PSF with estimates computed via RCA and CDL from images with a star density of 1 pixel per star.}
  \label{fig:psf_complex_sect}
\end{figure*}

\begin{figure*}[htpb]
  \centering
  \includegraphics[width=0.95\textwidth]{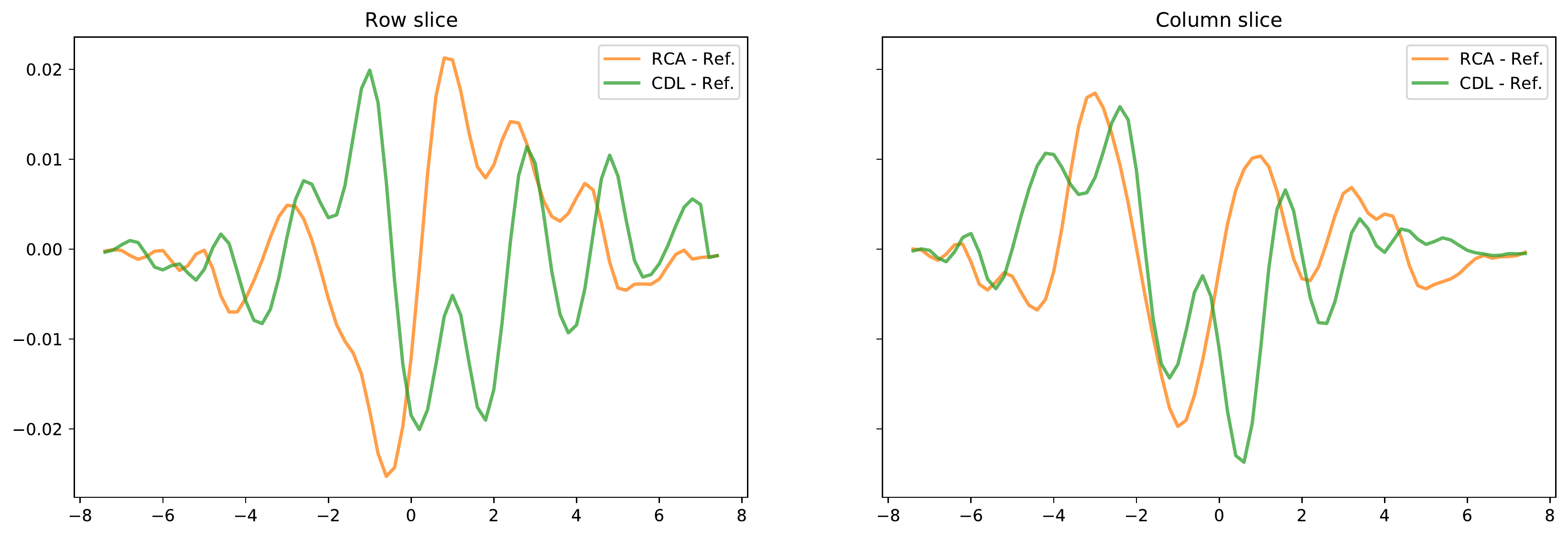}
  \caption{Row and column slices of the differences between the reference ``complex'' shape PSF and the estimates (a constant zero difference represents a perfect estimate) computed via RCA and CDL from images with a star density of 1 pixel per star.}
  \label{fig:psf_complex_secdif}
\end{figure*}

Selected examples from the performance comparison in~\sctn{perfcmp} of the main document are displayed in~\figs{psf_narrow_cntr}--\fign{psf_complex_secdif}.
The sub-pixel resolution PSF estimates shown in these figures were obtained by Lanczos interpolation of the pixel resolution PSFs estimated via RCA and CDL.

\end{document}